# A facile route to synthesize cubic gauche polymeric nitrogen

R.T. Chen[1] †, J. Zhang[1,2] †*, Z.L. Wang[1], K. Lu[1], Y. Peng[1], J.F. Zhao[1,2], X.D. Liu[3], S.M. Feng[1,2], R.B. Liu[3], C. Xiao[3] & C.Q. Jin*[1,2]

[1]Beijing National Laboratory for Condensed Matter Physics, Institute of Physics, Chinese Academy of Sciences, Beijing 100190, P. R. China.

[2]School of Physics, University of Chinese Academy of Sciences, Beijing 100190, P. R. China.

[3]School of Mechatronical Engineering, Beijing Institute of Technology, Beijing 100081, P. R. China

†These authors contributed equally.

*Corresponding author: zhang@iphy.ac.cn; jin@iphy.ac.cn

The polymerized nitrogen material with single bonds is considered one of the candidates for storing huge chemical energy due to the significant energy difference between triply bonded di-nitrogen and singly bonded nitrogen [1]. It was anticipated that the greatest utility of fully single-bonded nitrogen would yield a tenfold improvement in detonation pressure over HMX (one of the more powerful high explosives) [2] and therefore polymerized nitrogen is of great interest to physical sciences. It was predicted in 1985 that molecule nitrogen would polymerize to atomic solid at high pressure [3]. Later, cubic gauche structure (cg-N) was proposed [4]. It was not until 2004 that Eremets et al. [5] successfully prepared cg-N directly from molecular nitrogen at pressures exceeding 110 GPa and temperatures above 2000 K, using a laser-heated diamond cell [6]. Over the past few years, the plasma-enhanced chemical vapor deposition (PECVD) technique has been developed to leverage the non-equilibrium plasma environment for synthesizing polymerized nitrogen materials [7-12].

In this communication, the long-sought cg-N with N-N single bond has been synthesized for the first time by a thermal-driven-only chemical route at ambient conditions. The successful synthesis of cg-N was achieved by first creating a solution of azides, which was then pretreated under vacuum conditions. Following the pretreatment, the resultant concentrated azide was heated at temperatures ranging from 260°C to 330°C for a reaction time of 3 hours, ultimately leading to the formation of cg-N (Fig. S1 in Supplementary Materials). The emergent intense Raman peak characterized of cg-N provides solid evidence that the double bonded nitrogen-nitrogen transforms into a single bond form, which agrees well with cg-N structure. To date, this is the only work achieving the quantity of cg-N synthesized at ambient conditions by a facile route that can be further developed for the scalable synthesis and applications of polymerized nitrogen-based materials as high energy density materials.

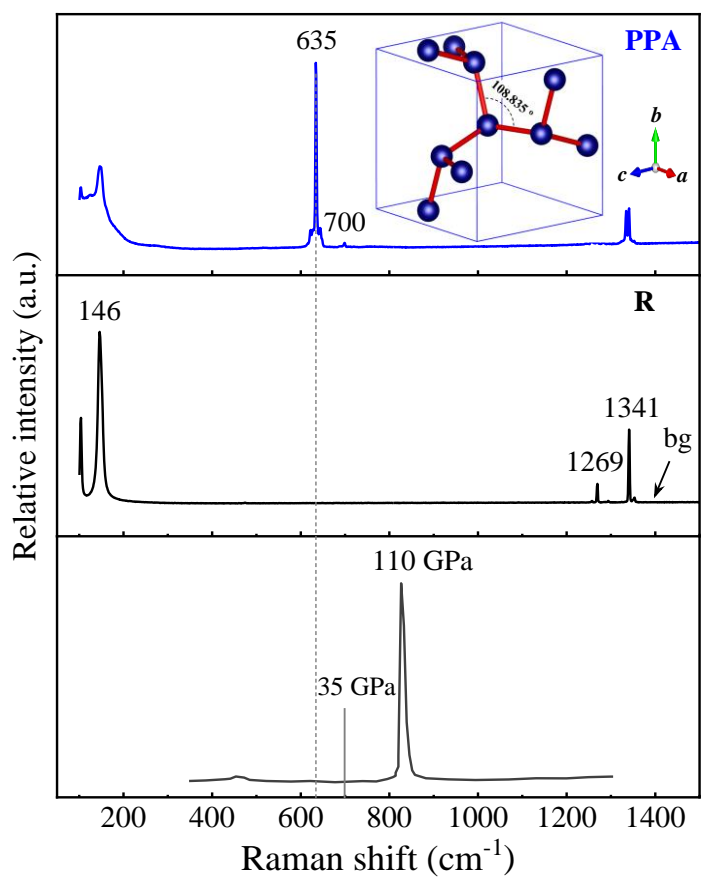

**Fig. 1** The Raman spectra: the black line and blue line are the Raman spectra for KN₃ raw material (R) and polymerized KN₃ (PPA), where the fingerprint peak indicative of cg-N is detected. For reference, the Raman spectrum of cg-N at 110 GPa extracted from experimental observation [5] and the calculated vibron position of cg-N from the A mode at 35 GPa [13] are shown in the bottom panel, which can be extrapolated linearly to ambient pressure as shown by the dashed line (see Fig. S3 in Supplementary Materials).

The inset of Fig. 1 shows the crystal structure of cg-N, in which nitrogen atom naturally adopts an sp² hybridization. Therefore, each N atom can be single-bonded to three nearest neighbors at equal distance and the remaining *p* orbital is occupied by the lone pair electrons. Thus, cg-N exhibits a tetrahedral structure similar to that of the ammonia (NH₃) molecule. The bond angle of N-N-N extracted from the compressed structure of cg-N is 108.835° [6] and it tends to increase as the structure recovers to ambient pressure [13]. Consequently, the N-N-N bond angle in cg-N slightly deviates from the ideal tetrahedral angle of 109.28°, resulting in minimal structural tension in cg-N. Furthermore, it is expected that cg-N will be stable at ambient pressure because the cg-structure is energetically the most favorable. This is well-demonstrated by theoretical work, which has shown that no imaginary frequencies are observed in its phonon spectrum [14].

Fig. 1 displays the Raman spectra of KN$_3$ samples subjected to various experimental processes. Primarily, it includes the Raman spectrum of the raw material KN$_3$ (R) and the Raman spectrum of polymerized KN$_3$ (PPA: polymerized potassium azide), which was obtained through a facile route reported in this work. For comparison, the Raman spectrum of heating-only KN$_3$ (UPA: unpolymerized potassium azide) is provided in the Fig. S2 in Supplementary Materials. As is well known, the Raman vibron peaks in azides principally originate from the N$_3^-$ anion. The Raman spectrum of cg-N dominated by the A mode was proposed theoretically by Caracas [13] and the position at 35 GPa is included in the bottom of Fig. 1 as a reference. Additionally, the experimental observation of cg-N at 110 GPa by Eremets et al. [5] is also shown. Based on these two references, a shift rate of 1.87 cm$^{-1}$/GPa can be calculated. By extrapolating this line to ambient conditions, it is estimated that the Raman peak position of cg-N should be around 635 cm$^{-1}$ (as shown in Fig. S3 in the Supplementary Materials), comparable to the calculated positions at ambient pressure condition [15]. The sharp peaks of KN$_3$ at 146, 1269 and 1341 cm$^{-1}$ are assigned to the vibrational lattice mode, the first overtone of the IR active bending $\nu 2$ mode and the symmetric stretching $\nu 1$ mode of the azide ion, respectively, which matches well with the free-standing KN$_3$ [11]. The Raman vibrons for the UPA sample are the same as those of R sample (see Fig S2 in Supplementary Materials), indicating that simply heating KN$_3$ can't lead to the formation of cg-N. As reported in recent work, plasma plays a dominant role in the cg-N synthesis, even when similar heating conditions were applied [11]. After pretreating and heating, the Raman spectrum of the PPA sample is highlighted in the blue line in Fig. 1, where the emergent intense line at 635 cm$^{-1}$ and the much weaker line at 700 cm$^{-1}$ are definitively observed. The line at 635 cm$^{-1}$ and 700 cm$^{-1}$ for polymerized potassium azide sample PPA are comparable to the theoretical results when extrapolated to ambient pressure, as the Raman modes tend to soften with decreasing pressure. It also aligns with the recent work by Xu et al. [11]. Consequently, the lines at 635 cm$^{-1}$ and 700 cm$^{-1}$ are assigned to the pore breathing A symmetry and N-N tilting T(TO) symmetry Raman-active modes of cg-N, respectively. Therefore, the new Raman peak at 635 cm$^{-1}$ is the fingerprint peak of cg-N, unequivocally indicating the successful synthesis of cg-N. The synthesis conditions of cg-N were systematically investigated while the optimized polymerized potassium azide sample PPA was obtained at 300 °C as shown in the Fig S1 in Supplementary Materials.

In order to provide a quantitative comparison, the intensity ratios of the characteristic Raman peaks are used as a measure of the abundance of the cg-N phase in the sample. The partial line intensities of KN₃ and cg-N are recorded in Table SI (Supplementary Materials). Here, the amount of cg-N synthesized is reflected by the peak intensity at 635 cm⁻¹, while the peak at 1341 cm⁻¹ represents the unreacted KN₃. Herein, a simplified conversion degree from KN₃ to cg-N is defined by the peak intensity ratio. The intensity ratios of KN₃ at 146 and 1341 cm⁻¹, calculated by $(I_{146}\text{-}I_{bg})/(I_{1341}\text{-}I_{bg})$, are 2.35, 1.81, and 2.20 for the KN₃ raw material (R), KN₃-heating only sample (UPA), and polymerized KN₃ sample (PPA), respectively, and they remain little changed. For samples of R and UPA, no signals of cg-N are found. After pretreating, for PPA sample the vibron intensity of cg-N enhances dramatically and the ratio of cg-N to KN₃ is 5.15. Therefore, it means that the pretreating process strongly promotes the formation of cg-N.

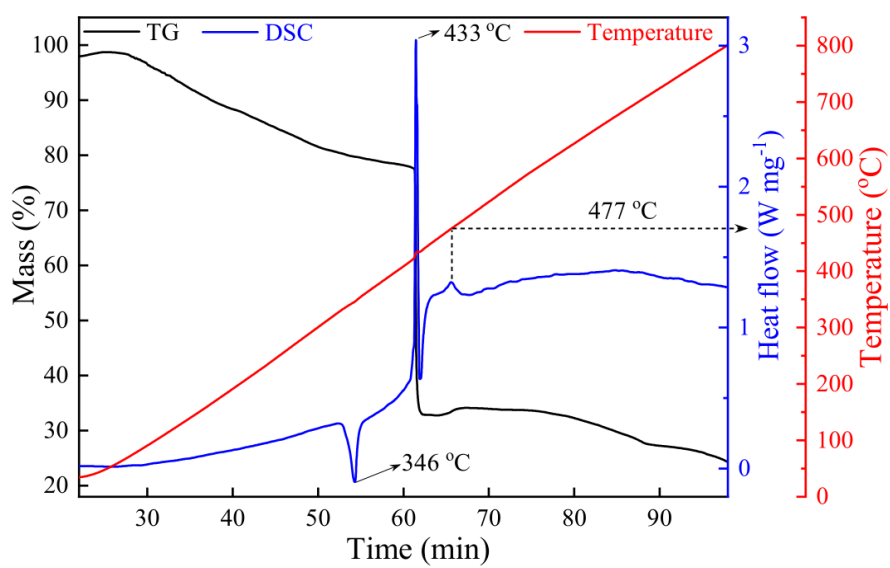

**Fig. 2 The TG-DSC curves of polymerized KN₃ sample (PPA). TG, DSC and temperature curves are shown in black, blue and red lines, respectively. The endothermic peak at 346 ºC and the exothermic peak at 433 ºC correspond to the melting and decomposition of remnant or unreacted KN₃ raw material in PPA sample, respectively. The new exothermic peak emerging at 477 ºC is from the decomposition of cg-N produced in PPA sample.**

To study the thermal stability of polymerized KN₃ (PPA) sample, TG-DSC measurement was performed. Fig. 2 displays the time dependence of mass loss, heat flow, and temperature changes during these measurements. The endothermic peak at 346 °C corresponds to the melting of unreacted KN₃ in the PPA sample, while the exothermic peak at 433 °C corresponds to its decomposition. This is further confirmed by the significant mass loss observed in the TG curve at 433 °C. Additionally, a new exothermic peak appears at 477 °C, which aligns well with the decomposition temperature of 488 °C reported for the cg-N sample in a recently published work [11]. However, the exact amount of cg-N in the PPA sample cannot be determined accurately due to baseline shifts in the TG curves. The shift was caused by instantaneous volume expansion or pressure increases resulting from the thermal decomposition of cg-N. Here, it was found that the thermal decomposition of unreacted $KN_3$ occurs between 410 to 500 ℃, highly depending on the preparation process of PPA samples after several measurements. That should be related to the discrepancies in sample microstructure.

The long-sought cg-N to be synthesized at ambient condition was successfully realized at ambient condition using a facile route. We got same polymeriized nitrogen upon replacing $KN_3$ by $NaN_3$. This should be applicable to other azides. It seems that the polymerization is regardless of the alkali metal in use. This is the simplest way so far to experimentally obtain cg-N.

**Acknowledgments**

The work is supported by NSF, MOST & CAS of China through research projects.